\documentclass[10pt]{myelsart}
\usepackage{amssymb,amsmath,float}
\usepackage{epsfig,graphics,subfigure}

\begin{document}
\begin{frontmatter}

\title{Condensate localization in a quasi-periodic structure}
\author{Y. Eksioglu},
\author{P. Vignolo\corauthref{cor1}} and
\author{M.P. Tosi}
\corauth[cor1]{Corresponding author, e-mail: {\tt vignolo@sns.it}}
\address{NEST-INFM and Scuola Normale Superiore,
Piazza dei Cavalieri 7, I-56126 Pisa, Italy}
\maketitle

\begin{abstract}
We propose a set-up of optical laser beams by which one may realize 
a quasi-one-dimensional Fibonacci array of potential wells for 
a Bose-Einstein condensate. We use a Bose-Hubbard tight-binding 
model to evaluate the transport of superfluid 
$^{87}$Rb atoms driven by a constant force through 
such an array. We show that the minigaps that are generated in the 
spectral density-of-states by the 
quasi-periodic disorder give rise to prominent localization effects, 
which can be observed by measuring the tunnel output of matter into 
vacuum as a function of the intensity of the applied force.
\end{abstract}
\begin{keyword}
Bose-Einstein condensates, transport properties, optical laser applications
\PACS{42.62.-b,\,03.75.Lm}
\end{keyword}

\end{frontmatter}
\newpage
\section{Introduction}
It is well known from solid-state physics that the introduction of 
quasi-periodic disorder in a system of particles on a lattice profoundly 
modifies the single-particle spectral density of states (DOS) and 
induces localization at certain energies. Here we discuss a way in which 
these effects are manifested in 
a superfluid Bose-Einstein condensate and how they may be experimentally 
demonstrated.
	
In more detail, we use a Bose-Hubbard tight-binding model for a 
superfluid assembly of $^{87}$Rb atoms to calculate the transport 
of matter driven by a constant force through a quasi-one-dimensional 
(1D) array of potential wells obeying the Fibonacci sequence. 
Parallel calculations are carried out for 
the current flowing under the same conditions through a perfectly 
periodic array. As in our previous 
studies of transport in a condensate \cite{Vignolo2003a,Eksioglu2004a}, 
our calculations use a Green's function method in which the 
array is reduced by a renormalization/decimation technique to a single 
dimer connected to incoming and outgoing leads \cite{Farchioni1996a}. 
The current is assessed from the output of matter tunnelling into vacuum. 
We show that the minigaps that are generated in the DOS by 
quasi-periodicity give rise to prominent localization effects, 
which are revealed in the output current as a function of the intensity 
of the applied force.
	
We then propose a method by which one may realize in the laboratory a 
quasi-one-dimensional array of potential wells obeying the  Fibonacci 
sequence for a Bose-Einstein condensate. We use the 
idea that a quasi-periodic array of dimensionality d can be created 
by suitably projecting a periodic 
array of dimensionality 2d onto a space of dimensionality 
d \cite{Fujiwara1990a}. In the case of present interest, 
we propose a set-up of four optical laser beams to create a square 
lattice and a suitably directed hollow beam to confine the atoms in 
a strip which is oriented relative to the lattice according to the golden 
ratio. Gravity acting on a tilted assembly may be used to generate a 
constant drive.

\section{The model and its numerical solution}
Our theoretical approach has been described in detail in our previous 
studies \cite{Vignolo2003a,Eksioglu2004a} and here we report 
only its essential points. The 1D Bose-Hubbard Hamiltonian for $N$ 
bosons distributed inside $n_s$ potential wells is
 \begin{equation}
H_I=\sum_{i=1}^{n_s} 
\left[E_i |\,i\rangle\langle i\,|+\gamma_i(|\,i\rangle
\langle i+1\,|+|\,i+1\rangle
\langle i\,|)\right]\,.
\label{Hamiltonian}
\end{equation} 
where $E_i$ and $\gamma_i$ are site energies and hopping energies, 
respectively. In a tight-binding scheme the 
condensate wave function $\phi_i(z)$ in a quasi-1D harmonic well can 
be represented by a Gaussian Wannier function of axial width 
$\sigma_i$ \cite{Slater1952a}. The parameters entering the effective 
Hamiltonian are then given by
 \begin{equation}
\hspace{-2.5cm}E_i=\int dz\,\phi_i(z)\left[
-\frac{\hbar^2\nabla^2}{2m}+U_i(z)+\frac{1}{2}g_{b}
|\phi_i(z)|^2-Fz+{\mathcal C}\right]\phi_i(z)
\label{siteenergy}
\end{equation} 
and
\begin{equation}
\hspace{-2cm}\gamma_i=\int dz\,\phi_i(z)
\left[
-\frac{\hbar^2\nabla^2}{2m}+U_i(z)+\frac{1}{2}g_{b}
|\phi_i(z)|^2+{\mathcal C}\right]
\phi_{i+1}(z).
\label{hopenergy}
\end{equation} 
In Eqs.~(\ref{siteenergy}) and (\ref{hopenergy}) $U_i$ is the external 
potential acting on the bosons in the $i$-th well, $F$ is a constant 
external force, $g_b$ is the effective 1D boson-boson coupling strength, 
and $\mathcal C$ is a constant accounting 
for transverse confinement. 
Nonlinear interaction effects enter the self-consistent determination of the 
widths $\sigma_i$, so that the condensate wave function $\phi_i(z)$  
and the parameters in the Hamiltonian $H_I$ also 
depend on the number of bosons in each well \cite{Vignolo2003a}. 
This approach is well justified in the case of weak 
boson-boson coupling as for a $^{87}$Rb condensate, which we study in 
this work.
	
In the Green's function method the calculation of transport by bosonic 
matter waves through the array of potential wells does not require an 
explicit solution of the Hamiltonian 
(\ref{Hamiltonian}) \cite{Farchioni1996a}. 
As already 
noted, the array is reduced to a single dimer to which an incoming lead 
and an outgoing lead are connected. The incoming lead injects particles 
in the zero-momentum state and the outgoing lead, in 
the case of a perfectly periodic array, extracts them by allowing tunnel 
into vacuum after acceleration by the constant external force up to the 
edge of the Brillouin zone. Since the presence of the constant 
force tilts the array (cf. Eq. (\ref{siteenergy})), the outgoing lead 
has to be connected to a well whose position is 
determined by the magnitude of $F$ according to the above criterion. 
The position of the outgoing lead for each value of $F$ is preserved in 
going from a periodic to a Fibonacci array. The steady-state 
transport coefficient is then inferred from the scattered wave function 
of the leads in the presence of the effective dimer.

\section{Density of states}
In order to understand the results for the transport coefficient in 
the two cases of a periodic and a quasi-periodic array, it is useful 
to exhibit first the DOS for the two cases. Its calculation has been 
carried out by recursive algorithms such as those presented in Refs.
\cite{Vignolo1999a,Farchioni2000a}.

\begin{figure}[H]
\centering{
\epsfig{file=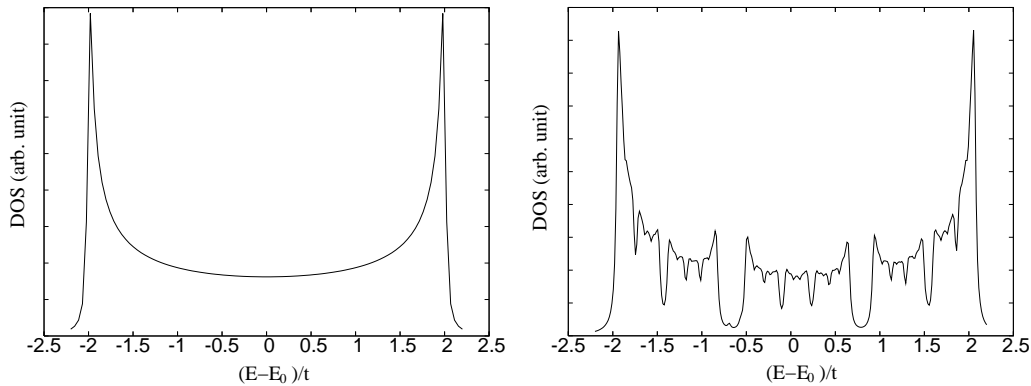,width=1.\linewidth}}
\caption{Total density of states DOS of a periodic array of potential 
wells (left panel) and of a quasi-periodic Fibonacci array (right panel), 
as a function of the energy $E$ referred to the band centre at 
energy $E_0$, for a total band width equal to $4t$.}
\label{fig1}
\end{figure}

The total DOS at energy $E$ is defined as
\begin{equation}
n(E)=-\frac{1}{\pi}{\rm{Im}}\sum_i^{n_s}\langle i\,
|\,\hat G(E)\,|\,i\rangle\,,
\end{equation}
where $\hat G(E)$  is the single-particle Green's function operator. 
Figure \ref{fig1} reports $n(E)$ as a function of 
energy over the whole energy band for a very long periodic array 
(1000 sites) and for the corresponding Fibonacci array. The latter
 has been generated by arranging two types of sites with 
energies $E_1$ and $E_2$ in a sequence determined according to the 
Fibonacci chain rule {\it ABBABABB.... }
The sequence is generated by the transformation rule $A\rightarrow B$   
and $B\rightarrow BA$.

It is evident from Fig. \ref{fig1} that the introduction of 
quasi-periodicity leads to a fragmentation of the 
spectrum of single-particle energies. This is a typical product 
of introducing disorder in a periodic 
system by quasi-periodic or even aperiodic modulations of the site energies. 
In particular, in the classical case of a quasi-periodic Fibonacci 
chain the spectrum is known to be a Cantor set with zero 
measure. The emergence of minigaps in the DOS causes the effects 
of particle localization that we shall illustrate in the next section.

\begin{figure}[H]
\centering{
\epsfig{file=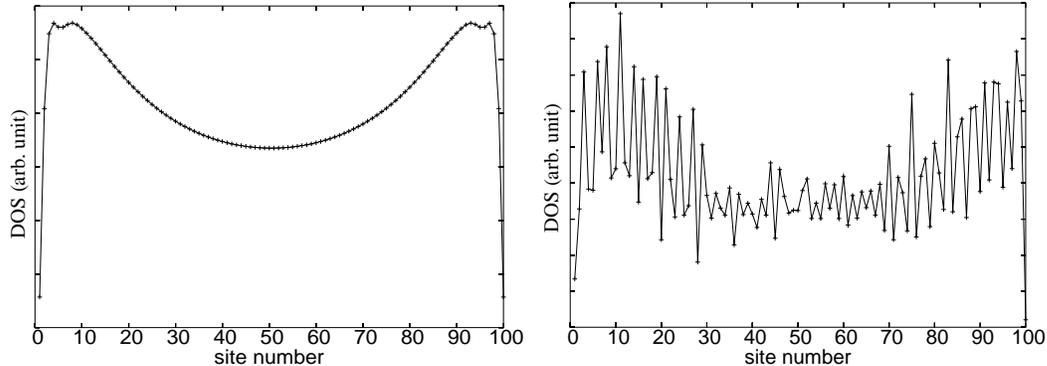,width=1.\linewidth}}
\caption{Projected DOS of a periodic array (left panel) and of 
a Fibonacci array (right panel), as a 
function of the site number. Both arrays extend over 100 sites.}
\label{fig2}
\end{figure}

Figure \ref{fig2} presents a comparison between the site-projected 
DOS of a periodic and a Fibonacci array of 100 wells under the action
of gravity. This quantity 
is defined as
\begin{equation}
n_i(E)=-\frac{1}{\pi}{\rm {Im}}\langle i\,
|\,\hat G(E)\,|\,i\rangle\,,
\end{equation}
and is shown in Fig. \ref{fig2} at the energy $E=E_0-2t$, that is 
at the bottom of the energy band. The site-projected DOS 
in the Fibonacci chain has a general envelope 
resembling that of the periodic chain, but 
is modifed by the quasi-periodicity favouring the population of 
a subset of sites.

\section{Localization from quasi-periodicity}
\label{localization}
\begin{figure}[H]
\centering{
\epsfig{file=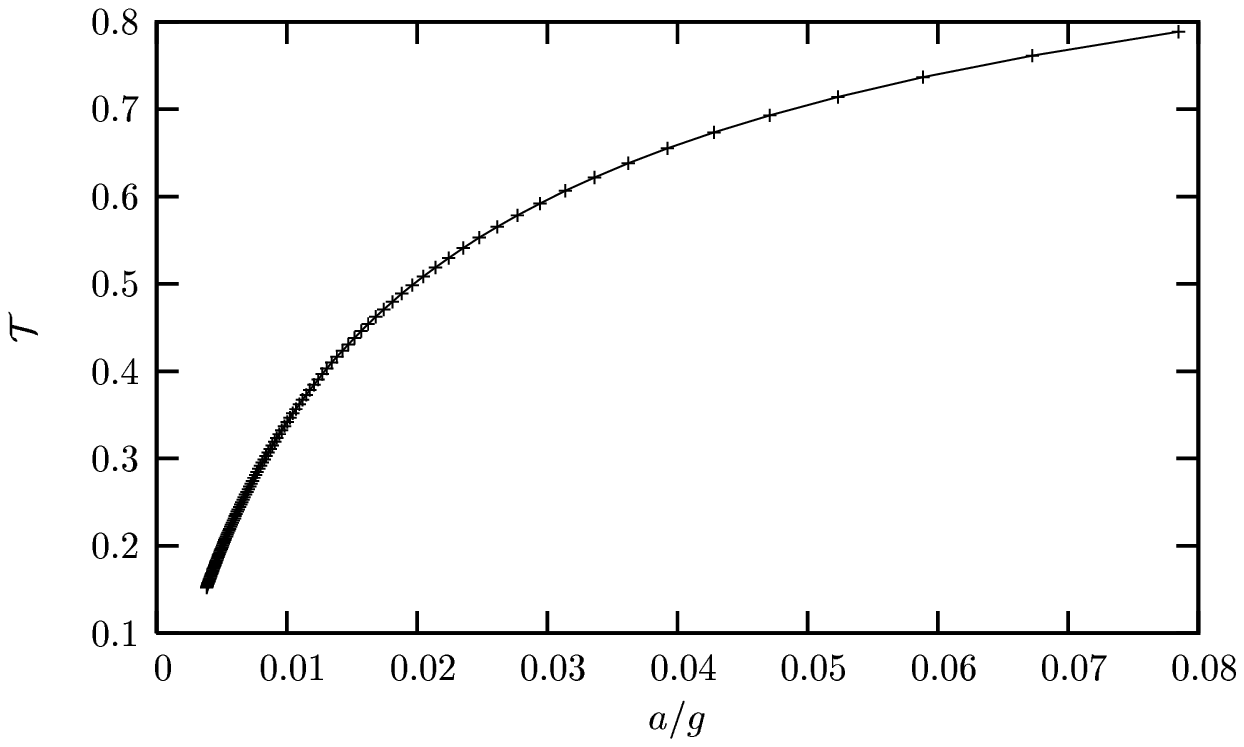,width=0.45\linewidth}
\epsfig{file=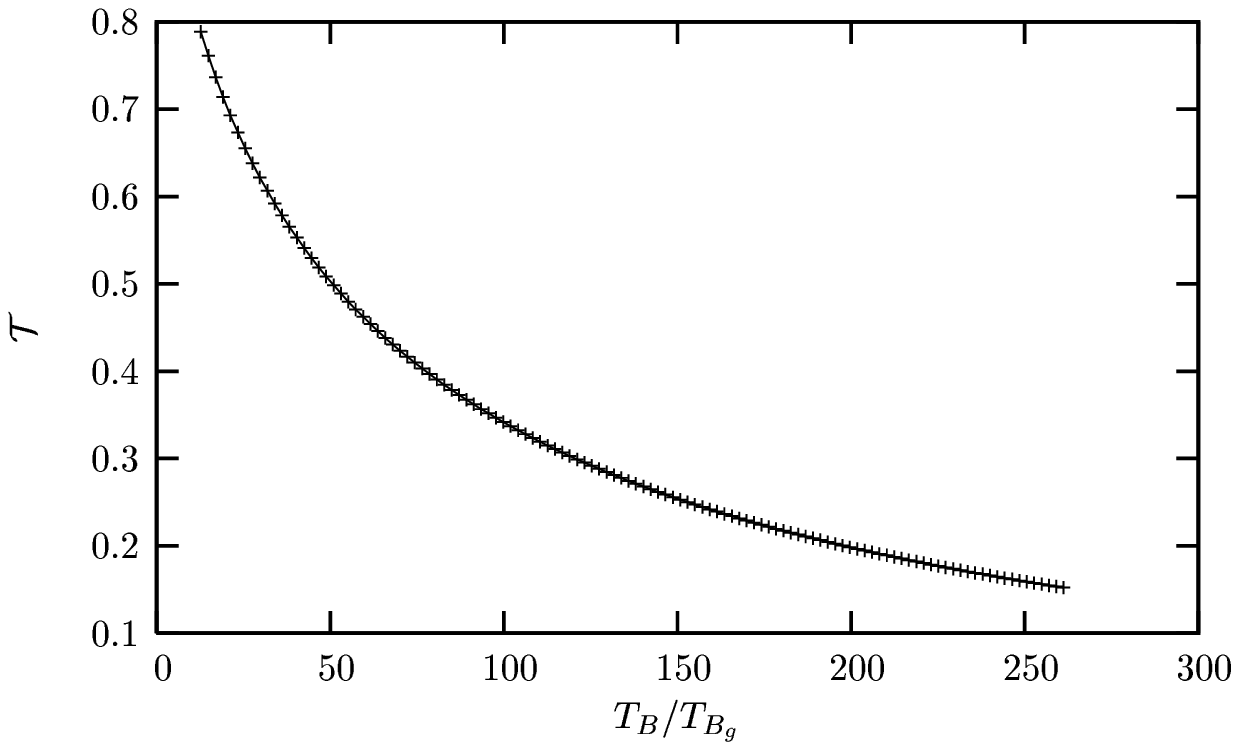,width=0.45\linewidth}}
\centering{
\epsfig{file=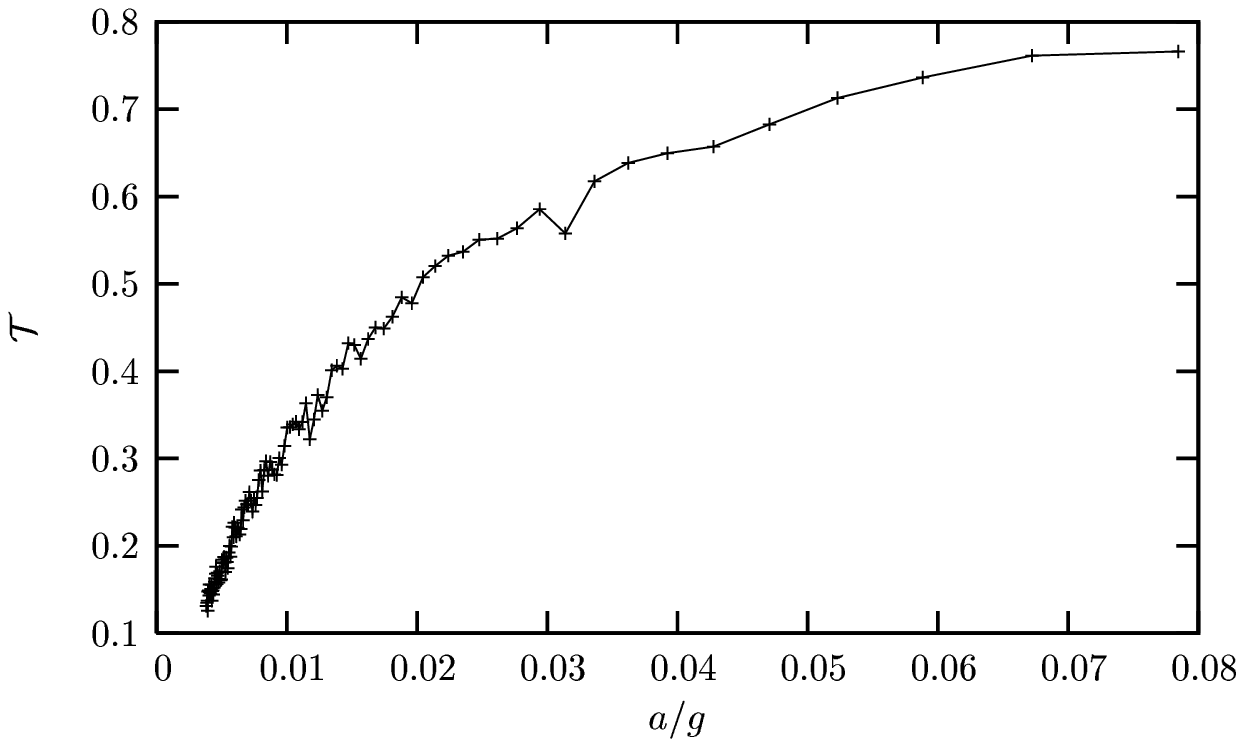,width=0.45\linewidth}
\epsfig{file=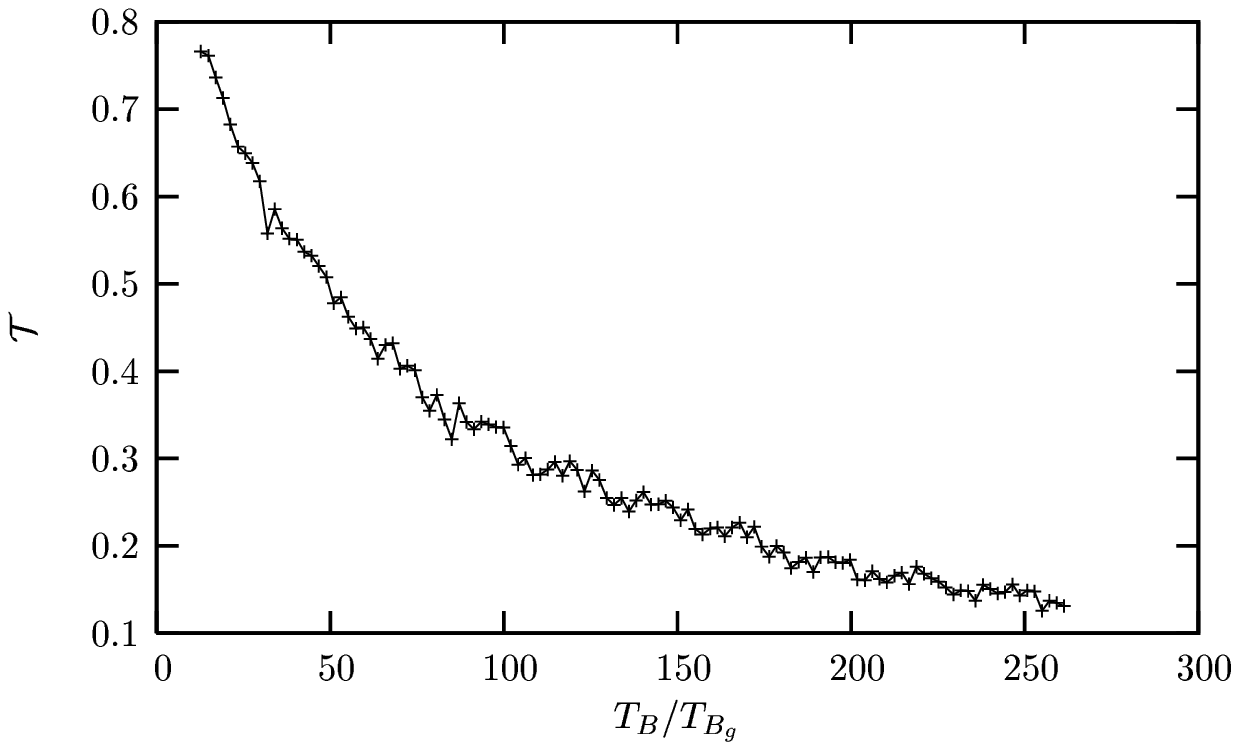,width=0.45\linewidth}}
\centering{
\epsfig{file=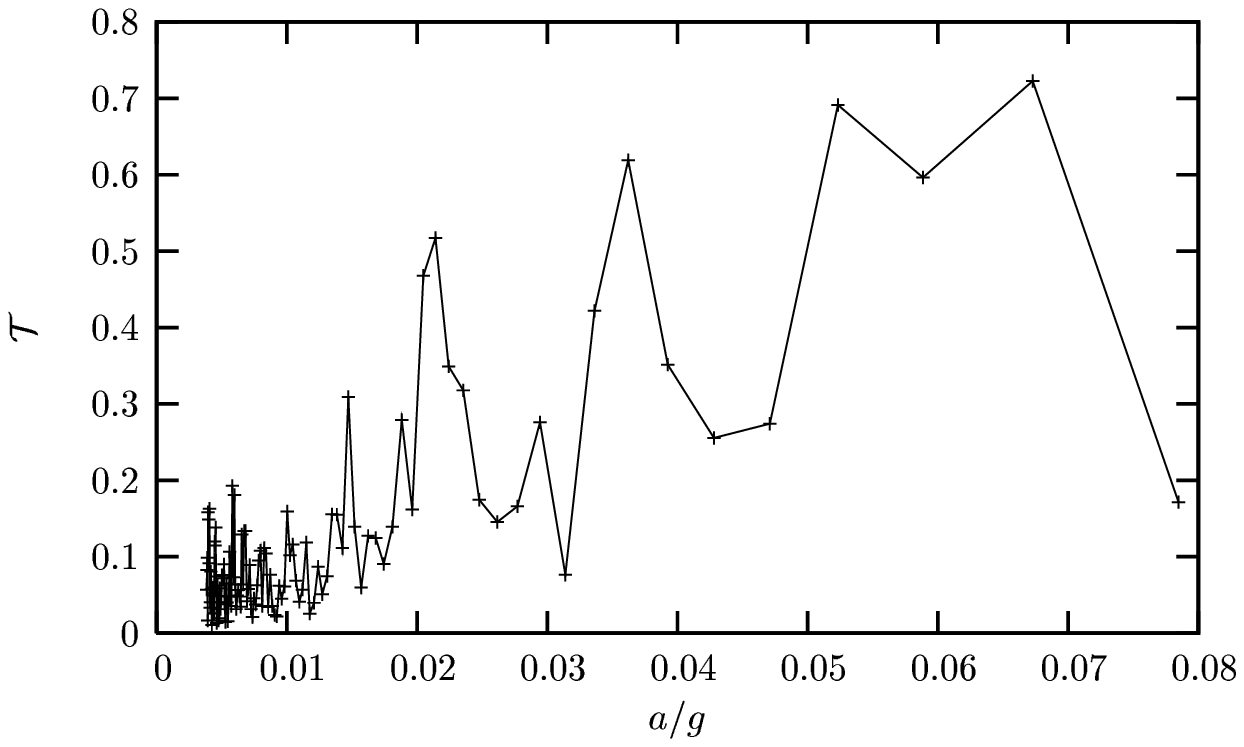,width=0.45\linewidth}
\epsfig{file=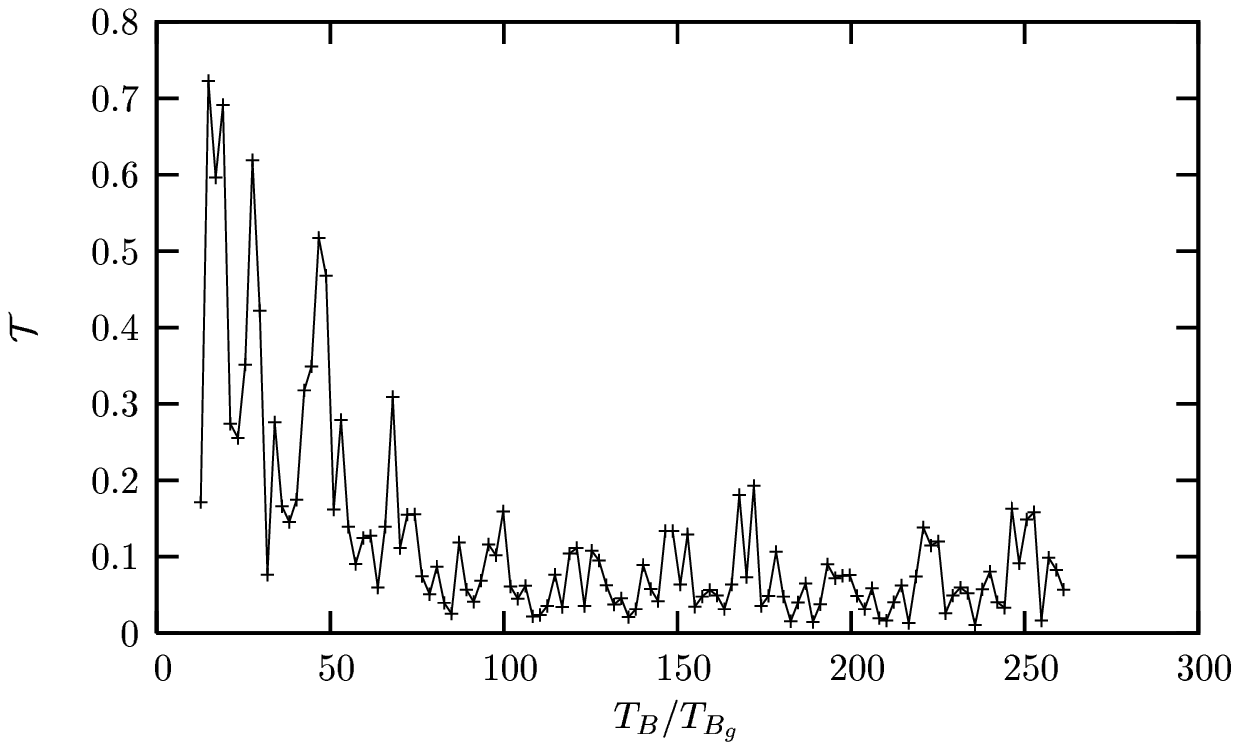,width=0.45\linewidth}}
\centering{
\epsfig{file=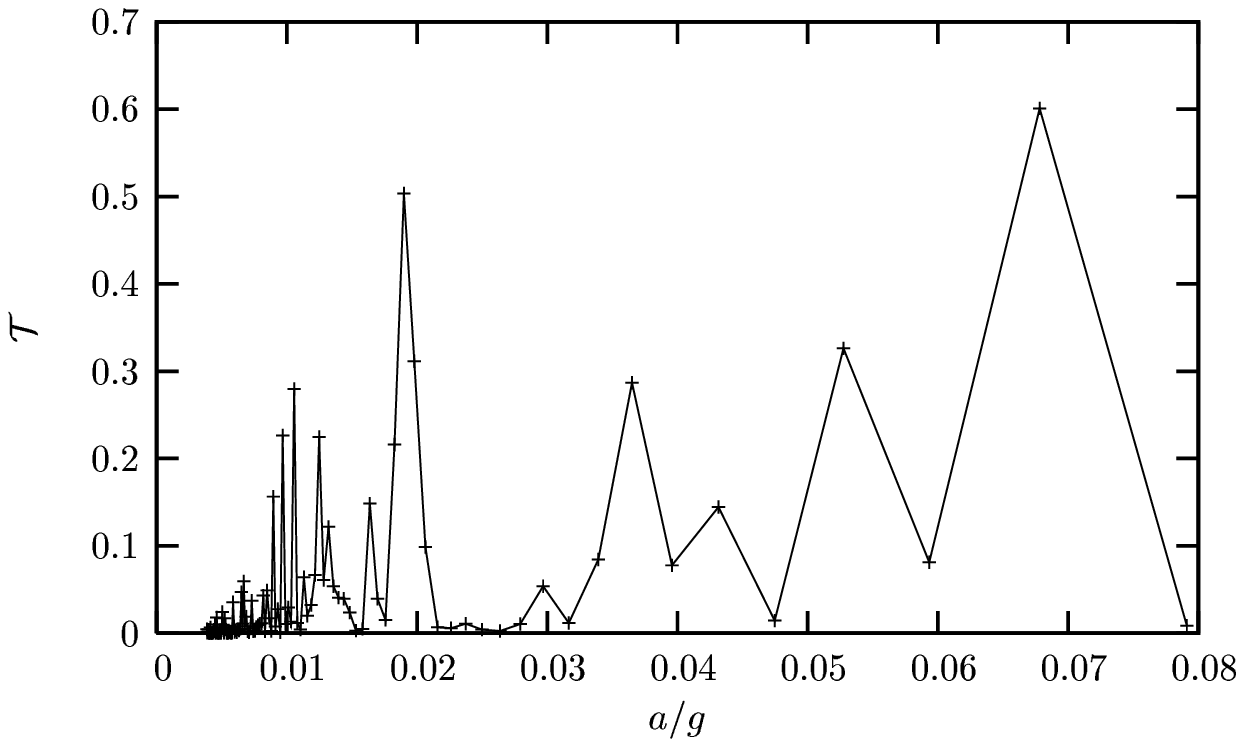,width=0.45\linewidth}
\epsfig{file=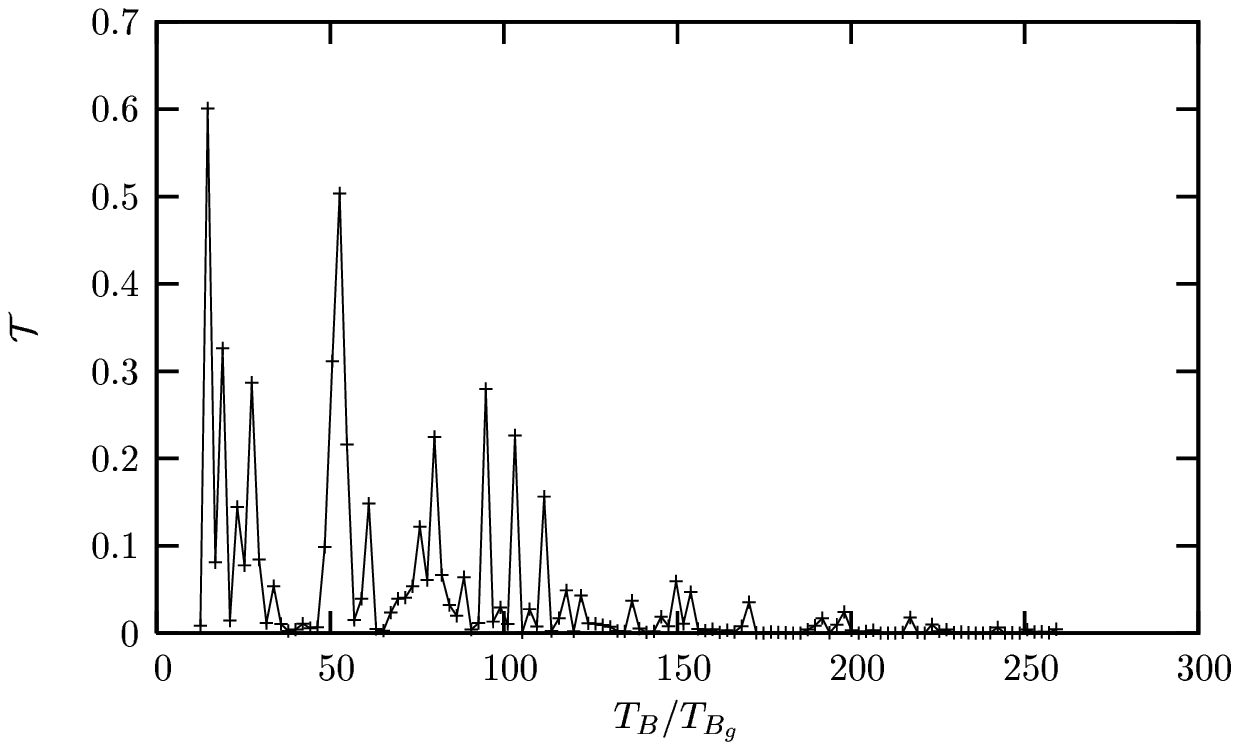,width=0.45\linewidth}}
\caption{Condensate transmittivivity as a function of its acceleration 
$a = F/m$ (in units of the acceleration 
of gravity $g$, first column) and of its inverse
$g/a=T_B/T_{B_g}$ for a periodic array (first row) and for Fibonacci 
arrays with increasing quasi-periodic disorder. The second, third, 
and fourth row correspond to
$\Delta E/\overline E=3\times 10^{-5}$, 
$3\times10^{-3}$, and $10^{-2}$, respectively.}
\label{fig3}
\end{figure}

The transmittivity coefficient $\mathcal{T}$ of condensate matter through 
the quasi-1D array is determined by the Green's function element 
describing the coherence between the input site and the output site 
(see Ref. \cite{Vignolo2003a} for a full discussion of the treatment 
and of the computational method). This coefficient depends 
on the magnitude of $F$ for each given value of the energy 
difference $\Delta E=|E_1-E_2|$. Figure \ref{fig3} reports 
our results for the case of a periodic array ($\Delta E=0$) and for 
three Fibonacci arrays constructed with 
increasing values of the ratio $\Delta E /\overline E= 2|E_1-E_2|/|E_1+E_2|$.
The transmittivity is plotted in Fig. \ref{fig3} as a 
function of the magnitude of the force $F$, expressed either through 
the ratio of the acceleration $a = F/m$ 
to the acceleration of gravity $g$ or through its inverse, that is the 
ratio of the corresponding periods of Bloch oscillations $T_B/T_{B_g}=g/a$. 
Notice that the lines drawn in Fig. \ref{fig3} serve the only purpose of 
guiding the eye from each data point to the next.

It is seen from Fig. \ref{fig3} that, while in the case of the periodic 
array the trasmittivity coefficient is a smoothly increasing (nonlinear) 
function of the drive, the introduction of even a modest amount of 
quasi-periodic disorder (for $\Delta E/\overline E = 3\times10^{-3}$, in the 
third panel from the top) suffices to introduce a 
great deal of structure. In fact, the transmittivity coefficient is 
directly proportional to the outgoing 
current, since the difference in chemical potential between the two leads 
is fixed by the total band-width and does not depend on the strength 
of the drive \cite{Vignolo2003a}. Thus the low values of $\mathcal{T}$  
in the third panel in Fig. \ref{fig3} directly reflect low values of the 
particle current due to scattering against the quasi-periodic 
disorder. The minima effectively become zeroes in the bottom panel, 
corresponding essentially to localization of the atoms in a finite array 
with quasi-periodic disorder of 1$\%$ magnitude.

\section{Proposed set-up for an atomic Fibonacci waveguide}
Although our numerical illustration has referred to the situation in 
which the quasi-periodic disorder is created by modifying the site 
energies, it may be easier to realize experimentally a set-up in 
which the disorder is introduced in the hopping energies. 
The behaviour of matter waves propagating through such a quasi-periodic 
array should not differ qualitatively from that illustrated in 
Sec. \ref{localization} above.

\begin{figure}[H]
\begin{center}
\epsfig{file=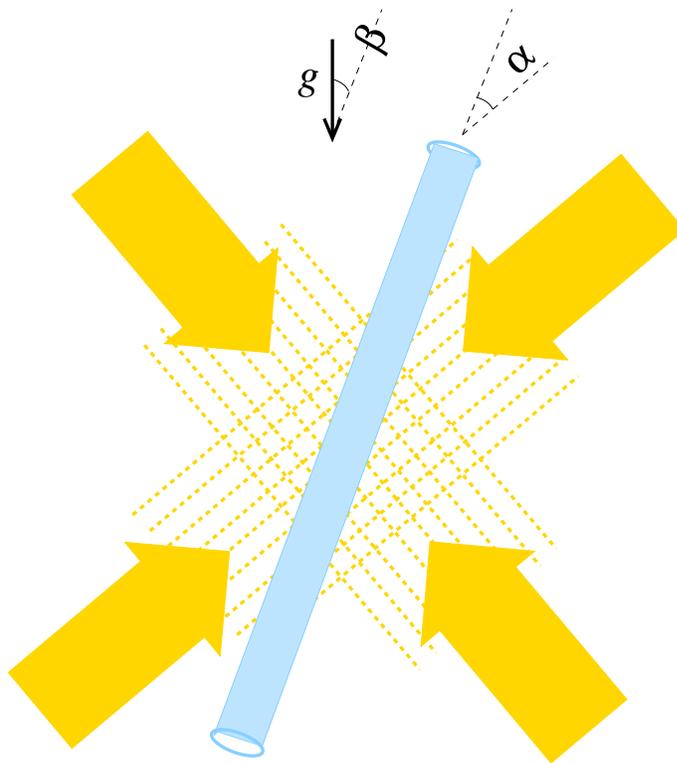,width=0.65\linewidth}
\end{center}
\caption{Schematic representation of a five laser-beam configuration 
to create a quasi-one-dimensional Fibonacci array of potential wells 
with quasi-periodic hopping energy. Four beams generate a square 
optical lattice and a hollow beam confines the condensate to a strip 
with slope $\alpha=\arctan(2/(\sqrt{5}+1))$ relative to an axis of 
the lattice. The angle $\beta$ between the hollow beam and the 
vertical direction determines the driving force as $F=mg\cos\beta$.}
\label{fig4}
\end{figure}

A schematic drawing of the set-up of optical lasers that would create 
an atomic Fibonacci wave guide is shown in Fig. \ref{fig4}. 
Here, two pairs of counter-propagating laser beams create a square optical 
lattice. The projection of this lattice on a line at an angle 
$\alpha=\arctan(2/(\sqrt{5}+1))$  relative to the lattice 
creates a quasi-periodic sequence of bond lengths, and hence of 
hopping energies, which obey the Fibonacci rule. The atoms can be made to travel along the sequence by pointing along 
this direction a hollow beam
(for a description of the latter see for example the work 
of Xu {\it et al.} \cite{Xu2001a}).
	
As for the constant external drive, a simple way to create and control 
it would be to tilt the whole set-up by an angle $\beta$ relative to the 
vertical axis (see Fig. \ref{fig4}). In this case the external force acting
on the condensate atoms is $F=mg\cos\beta$.

\section{Conclusions}
In summary, we have shown that localization can result in a Bose-Einstein 
condensate propagating along a quasi-periodic array of potential wells 
from the opening of sharp depressions ("minigaps") in 
the spectral density of states due to quasi-periodic disorder in the 
site energies. A similar situation will arise when the quasi-periodic 
disorder is generated in the hopping energies for condensate atoms from 
well to well in an array, and we have proposed a five-laser set-up by 
which this situation could be created in the laboratory.

It also seems worthwhile to recall that in Ref. \cite{Eksioglu2004a} 
we have interpreted the structure in the transmittivity coefficient 
as being the result of interference between matter waves propagating through 
the quasi-periodic array. This interpretation was based on the qualitative 
affinity between the minigaps in the DOS of the quasi-periodic array in 
Fig. \ref{fig1} and the minigap that can be created at the centre of the 
band in a periodic system by doubling its periodicity. 
The doubled-period set-up has an optical 
analogue in  an apparatus that performs beam splitting followed by
 beam interference, and leads to a (regular) structure of maxima and 
minima in the transmittivity coefficient as a function of a constant 
drive.

\ack{This work has been partially supported by Scuola Normale Superiore 
di Pisa through an Advanced Research Initiative and by the Istituto 
Nazionale di Fisica della Materia through the PRA-Photonmatter Programme.}


\begin{thebibliography}{10}
\expandafter\ifx\csname url\endcsname\relax
  \def\url#1{\texttt{#1}}\fi
\expandafter\ifx\csname urlprefix\endcsname\relax\def\urlprefix{URL }\fi

\bibitem{Vignolo2003a}
Vignolo P., Akdeniz Z., and Tosi, M.P., 2003, J. Phys. B, 36, 4535.
\bibitem{Eksioglu2004a}
Eksioglu, Y., Vignolo, P., and Tosi, M.P., 2004, Optics Commun., in press.
\bibitem{Farchioni1996a}
Farchioni, R. Grosso, G., and Pastori Parravicini, G., 1996, 
Phys. Rev. B, 53, 4294.
\bibitem{Fujiwara1990a}
Fujiwara, T. and Ogawa, T., 1990, Quasicrystals (Springer, Berlin).
\bibitem{Slater1952a}
Slater, J.C., 1952, Phys. Rev., 87, 807.
\bibitem{Vignolo1999a}
Vignolo, P., Farchioni, R., and Gross, G., 1999, Phys. Rev. B, 59, 1065.
\bibitem{Farchioni2000a}
Farchioni, R., Gross, G., and Vignolo, P., 2000, Phys. Rev. B, 62, 12565.
\bibitem{Xu2001a}
Xu, X., Kim, K., Jhe, W., and Kwon, N., 2001, Phys. Rev. A, 63, 063401.

\end{thebibliography}
\end{document}